\documentclass{jpsj-suppl}
\usepackage{times}
\usepackage{color}
\usepackage{subfigure}
\usepackage{here}
\title{Contribution of Berry Curvature to Thermoelectric Effects}

\author{Yo Pierre Mizuta,   Graduate School of Natural Science and
Tecnology, Kanazawa University \thanks{E-mail address: mizuta@cphys.s.kanazawa-u.ac.jp} 
\newline 
Fumiyuki Ishii,   Faculty of Mathematics and Physics, Kanazawa University}
\inst{Kakuma, Kanazawa, Ishikawa 920-1192 
}
\recdate{September 30, 2013}
\abst{Within the semiclassical Boltzmann transport theory, the formula
for Seebeck coefficient $S$ is derived for an isotropic two-dimensional
electron gas (2DEG) system that exhibits
anomalous Hall effect (AHE) and anomalous Nernst effect (ANE) originating
from Berry curvature on their bands. Deviation of $S$ from the value
$S_0$ estimated neglecting Berry curvarture is computed for a
special case of 2DEG with Zeeman and Rashba terms. The result shows
that, under
certain conditions the contribution of Berry curvature to Seebeck
 effect could be non-negligible. Further study is needed to clarify the effect of additional contributions
 from mechanisms of AHE and ANE other than pure Berry curvature.}

\kword{thermoelectric effect, anomalous Hall effect, anomalous Nernst
effect, Berry curvature, two-dimensional electron gas}

\begin{document}
\maketitle

\section{Introduction}
Thermoelectric effects, the mutual conversion of heat and electricity in
materials, have been attracting much attention these days
due to their great potential for applications such as eco-friendly power generation
from waste heat, or Peltier cooler for small devices.  Much efforts have
been made to find materials with better efficiency of conversion
achieved by larger value of $Z=\sigma S^2 /\kappa$, where $\sigma,\ S,\
\kappa$ are longitudinal electrical conductivity, Seebeck coefficient, and
thermal conductivity, respectively.  We now have, based on those works, some
guides in the quest for better materials, among which are the low
dimensionality of the system,
expected to make $\sigma$ large while keeping the magnitude of $S$~\cite{hickseffect1993}, or some specific structures which are believed to suppress thermal conduction efficiently. 

What we present in this paper is another ingredient, namely the  transverse
components  $\sigma_{xy}$ and $\alpha_{xy}$  of the tensors $\tilde{\sigma},\ \tilde{\alpha}$ in the expression of charge current ${\bf
j}=\tilde{\sigma}{\bf E}+\tilde{\alpha}(-\nabla T)$, where ${\bf E}$ and  $\nabla T$ are electric field and temperature gradient, respectively.
These transverse components naturally participate in the determination of
longitudinal quantity $S$ as will be seen in the formula  in the following section. As to the origin of $\sigma_{xy}$ and
$\alpha_{xy}$, one can think of either  external magnetic field or  spontaneous magnetization. 
 An example of the former is seen in ~\cite{jonsonthermoelectric1984}, where
two-dimensional electron gas (2DEG) in a quantizing magnetic field was 
studied. In what follows our target is the latter. These effects, the
appearance of $\sigma_{xy}$ and $\alpha_{xy}$ in zero magnetic field, are called anomalous Hall
effect (AHE) ~\cite{nagaosaanomalous2010} and anomalous Nernst effect (ANE) ~\cite{xiaoberry2010}, respectively. Their origins
had been a controversial problem for many decades, but it was
shown both experimentally ~\cite{leedissipationless2004,leeanomalous2004} and theoretically ~\cite{jungwirthanomalous2002,
onodaintrinsic2006,yaotheoretical2007,onodaquantum2008} that for some cases
so-called intrinsic contribution becomes important. It stems purely from some 
band structures and in the clean limit, the conductivities are expressed, as we will review in the following section,
as a functional of Berry curvature ${\bf \Omega}({\bf k})\equiv i
\langle \partial_{\bf k}u |\times |\partial_{\bf k}u \rangle$, where
$|u_{\bf k} \rangle$ is the periodic part of a Bloch state. Various topics related to Berry curvature are reviewed in Ref.~\cite{xiaoberry2010}. 
The target of the present study is to roughly estimate how much this intrinsic contribution
could affect the value of Seebeck coefficient in the case of 2DEG, the systems promising for thermoelectric
applications due to its low dimensionality.             

\section{Derivation of the formula for Seebeck coefficient  }
We derive below the semiclassical formula for Seebeck coefficient.
The starting point is the expression for charge current ${\bf j}$
obtained in Ref.~\cite{xiaoberry-phase2006}, which reads, 
\begin{equation}
{\bf j}=e\int \frac{d{\bf k}}{(2\pi)^2}g({\bf r},{\bf k}){\bf v}_{\bf k}+e\nabla_{\bf
 r}\times k_{\rm B}T({\bf r})\int \frac{d{\bf k}}{(2\pi)^2}{\bf \Omega}({\bf
 k})\log \left[1+\exp \left(-\frac{\varepsilon_{\bf k}-\mu}{k_{\rm B}T({\bf r})}\right) \right] \label{anomj},
\end{equation}
where  $e,\ g({\bf r},{\bf k)},\ T({\bf r}),\ {\bf v}_{\bf k},\ \varepsilon_{\bf k},\ {\rm 
and}\  \mu$ stand for the electron's charge($e<0$),  distribution function,  local temperature, velocity and energy of an electron with wave number {\bf k} and chemical potential, respectively. 
Hereafter we  use atomic units ($\hbar=1,\ |e|=1,\ m=1$) unless explicitly specified.

The second term is a correction that appears when spatial inhomogeneity (i.e., $T({\bf
r})$ in the present case) exists. 
Simplifying the second term following Ref.~\cite{xiaoberry-phase2006} and substituting the equation of motion of a perturbed (by {\bf E} field)
Bloch electron
\begin{equation}
{\bf v}_{\bf k}=\frac{\partial \varepsilon_{\bf k}}{\partial {\bf k}}-\dot{\bf
 k}\times{\bf \Omega}({\bf k}), \label{anomv}
\end{equation}
which was derived in Ref.~\cite{sundaramwave-packet1999}, for ${\bf v}_{\bf k}$ and the form of distribution
\begin{equation}
g({\bf r},{\bf k})=\tau_{\bf k}{\bf v}_{\bf k}\cdot \left[e{\bf
 E}+(\varepsilon_{\bf k}-\mu)\left(-\frac{\nabla T}{T}\right)
 \left(-\frac{\partial f}{\partial \varepsilon}\right)\right],  \label{g}
\end{equation}
obtained as the solution of Boltzmann transport equation within relaxation
time ($\tau_{\bf k}$) approximation, for $g({\bf r},{\bf k})$, we obtain
\begin{equation}
\left\{\begin{array}{l}
{\displaystyle \sigma_{xx}=\frac{e^2\tau}{2} \int d\varepsilon
 D(\varepsilon)v_0(\varepsilon)^2\left(-\frac{\partial f}{\partial
			\varepsilon}\right)=\sigma_{yy},} \vspace{2mm} \\
{\displaystyle \sigma_{xy}=-e^2\int d\varepsilon
 D(\varepsilon)\Omega_z({\bf k})=-\sigma_{yx},} \vspace{2mm} \\
{\displaystyle \alpha_{ij}=\frac{1}{e}\int d\varepsilon
\sigma_{ij}(\varepsilon)_{T=0}\frac{\varepsilon-\mu}{T}\left(-\frac{\partial
						f}{\partial
						\varepsilon}\right) }  \ \ \ \   {\rm for}\ \ \ \   i=x \ \ {\rm or}\ \  y,\  \ j=x\ \ {\rm or}\ \ y,
\end{array}
\right. \label{sigalp}
\end{equation}
where $D(\varepsilon)$ and 
$v_0(\varepsilon)\equiv d\varepsilon(k)/dk$ are density of states and  electron's group
velocity, respectively.
We restricted ourselves to the case of isotropic 2DEG, i.e.,
$\varepsilon({\bf k})=\varepsilon(k),\ {\bf \Omega}=(0,0,\Omega_z)$ and  assumed constant relaxation
time ($\tau_{\bf k}=\tau$).  

The formula for thermoelectric cofficients follow immediately by
substituting Eq. (\ref{sigalp}) for
$\tilde{\sigma}=[\sigma_{ij}]$ and $\tilde{\alpha}=[\alpha_{ij}]$ in the linear response relation ${\bf
j}=\tilde{\sigma}{\bf E}+\tilde{\alpha}(-\nabla T)$.  The result is,
\begin{equation} \left\{
 \begin{array}{l}
{\displaystyle S\equiv S_{ii} \equiv \frac{E_i}{\partial_i T}=\frac{\alpha + \beta \gamma}{1+\beta^2}}
 \vspace{2mm}\ \  \ \ \  {\rm for}\ \ \ \  i=x \ \ {\rm or}\ \  y \\
{\displaystyle N \equiv S_{xy} \equiv \frac{E_x}{\partial_y
 T}=\frac{\gamma - \alpha \beta}{1+ \beta^2}=-S_{yx}.} \label{SN}
\end{array} \right.
\end{equation}
Here we defined $\alpha \equiv \alpha_{xx}/\sigma_{xx}, \beta \equiv
\sigma_{xy}/\sigma_{xx},\ \gamma \equiv \alpha_{xy}/\sigma_{xx}$ for a simpler notation.

Note that the Seebeck coefficient $S_0$ estimated without considering Berry
curvature is obtained by setting $\beta=0$ and $\gamma=0$ in
Eq. (\ref{SN}), i.e. $S_0=\alpha$.

\section{Model}
Next we  estimate how much $S$ could deviate from $S_0$  according to Eq.(\ref{SN}).
We choose here, as an example system,  Zeeman-Rashba 2DEG
(ZR2DEG) most simply described by the following Hamiltonian:
\begin{equation}
 H=\frac{k^2}{2m^*}+\lambda({\bf e}_z\times {\bf k})\cdot \boldsymbol{\sigma}-\Delta \sigma_z
\equiv \frac{k^2}{2m^*}+{\bf h}({\bf k})\cdot \boldsymbol{\sigma}, \label{H}
\end{equation}
where $m^*,\ \lambda,\ {\bf e}_z,\ \boldsymbol{\sigma}, \ \Delta$, and ${\bf h}({\bf k}) \equiv (-\lambda k_y,\ \lambda k_x,\ -\Delta)$  are the effective mass of an electron, Rashba parameter, unit vector normal to the 2DEG plane (which we take to be $xy$-plane), spin operator,  exchange field for an electron, and effective "magnetic field", respectively.  Rashba term, the second one, is a spin-orbit coupling originating from the structural inversion asymmetry, which
 is present on the surface or at the interface of layers~\cite{bychkovproperties1984}.  The parameter
 $\lambda$ is determined not only by the kind of material but is
tunable with an applied gate-voltage~\cite{culceranomalous2003,winklerspin-orbit2003}.  Zeeman
 term, the last one, is assumed here to represent an exchange
 field created by localized spins at magnetic impurities and felt by a spin of Bloch electron.  Thus we here regard the Hamiltonian above
 as a model of 2DEG, induced for example at the interface of a 
 semiconductor heterostructure doped with magnetic impurities such as Mn.  Note, however, that  Eq.({\ref{H}) has  been studied not only as a model for 2DEG but also as a minimum model of 2D partial structure of 3D ferromagnetic metals accompanying Berry curvature ~\cite{onodaintrinsic2006,onodaquantum2008, nagaosaanomalous2010,xiaoberry2010}. 
This system has  eigenstates
\begin{equation}
|u_-\rangle=\left(\begin{array}{c} \sin(\theta/2)e^{-i\phi}
		\vspace{2mm} \\ 
-\cos(\theta/2) \end{array} \right), \ \ 
|u_+\rangle =\left(\begin{array}{c} \cos(\theta/2)e^{-i\phi}
	      \vspace{2mm} \\ \sin(\theta/2) \end{array} \right),
\label{eigv}
\end{equation}
where the angles $\theta=\theta({\bf k})$ and  $\phi=\phi({\bf k})$ are those of polar coordinates of
the vector ${\bf h}({\bf k})$, and the corresponding eigenenergies are
\begin{equation}
\varepsilon_{\pm}=\frac{k^2}{2m*}\pm \sqrt{\lambda^2k^2+\Delta^2}. \label{eng}
\end{equation}
Each band has the Berry curvature
\begin{equation}
 {\bf \Omega}({\bf k})_{\pm}=\left(0,\ 0,\
  \mp\frac{\lambda^2\Delta}{2(\lambda^2k^2+\Delta^2)^{3/2}}\right). \label{Omeg_model}
\end{equation}
Note that, in our model,  the absence of either term, Zeeman (time-reversal symmetry (TRS) breaking) or Rashba (inversion symmetry (IS) breaking), leads to zero Berry curvature, although in general, finite curvature is allowed in the presence of either TRS or IS. 

We plot Eq.({\ref{eng}) and Eq.(\ref{Omeg_model}) in Fig.(\ref{eng_Omeg}) for
some values of parameters.  We use hereafter $\eta \equiv m^* \lambda^2/\Delta$
instead of $\lambda$, since $\eta$ nicely controls the shape of
$\varepsilon_-$ band (i.e., $\eta=1$ is a critical value between
$\varepsilon_-$ bands with topologically different Fermi surfaces) and  in addition, all conductivities we need become independent of $m^*$ once we fix the values of the pair $\eta$ and $\Delta$. 

According to Ref.~\cite{culceranomalous2003} the value of $\lambda$ ranges from at most 0.69 meV$\cdot$nm for GaAs to at least 100 meV$\cdot$nm for HgMnTe.  For $\Delta$,  example values of  25 meV and 122 meV for (Ga,Mn)As are found in Ref.~\cite{jungwirthanomalous2002}.
 
%
 \begin{center}
 \begin{figure}[H]
  \subfigure[$\eta=0.1$]{%
  \includegraphics[width=5cm]{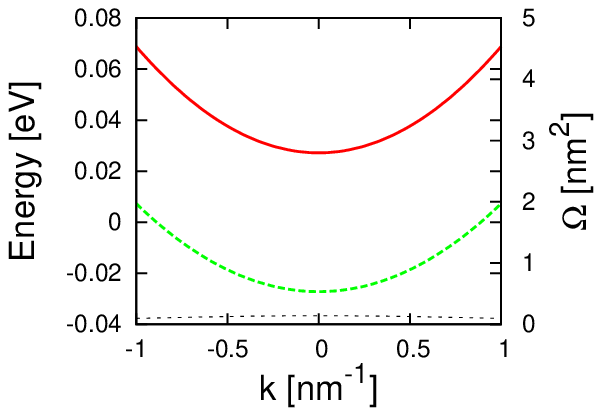}}%
  \subfigure[$\eta=1$]{%
  \includegraphics[width=5cm]{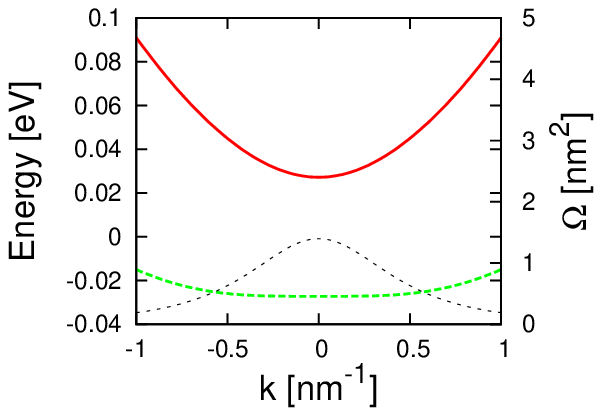}}%
  \subfigure[$\eta=3$]{%
  \includegraphics[width=5cm]{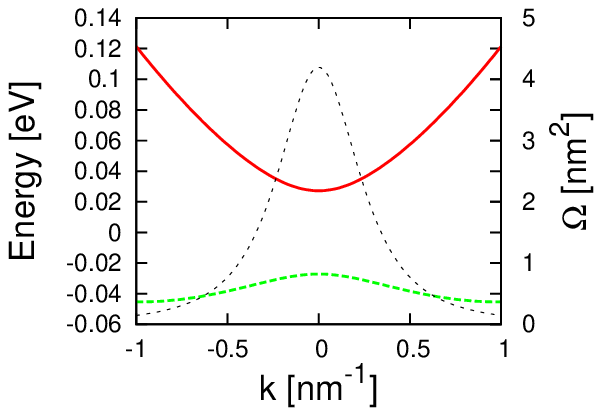}}%
   \caption{(Color online) Band dispersions(left axis, red:$\varepsilon_+$,
  green:$\varepsilon_-$) and Berry curvature(right axis, black broken) of ZR2DEG
  plotted as a function of wave number $k=|{\bf k}|$.  $m^*=1,\
  \Delta=10^{-3}$}
\label{eng_Omeg}
\end{figure}
 \end{center}

\section{Results and Discussions}
 Seebeck coefficient was calculated numerically for
 wide range of parameters using  Eq.(\ref{SN}). 
First of all, the relative deviation of $S$, the Berry curvature-included value, from conventionally estimated $S_0$, evaluated
by $r \equiv (S-S_0)/S_0$, becomes relatively large when (i) band parameter $\eta$ is of order 1 to 10, and 
(ii) chemical potential $\mu$ is situated near the $\varepsilon_-$ band edge. 

These trends (i) and (ii) can be understood qualitatively as follows: (i) $\eta \simeq 1$ means large density of states near the $\varepsilon_-$ band edge, where $\Omega(k)$ has its peak, leading to the grow of $\sigma_{xy}$ and $\alpha_{xy}$.  (ii) Low $\mu$ suppresses $\sigma_{xx}$ and thus make $\beta$ and $\gamma$ large. 

We therefore concentrate on some points in such parameter region and show the temperature
dependence and relaxation time dependence there in Fig.(\ref{S-T}).
\begin{center}
 \begin{figure}
  \subfigure[$\eta=1,\ \Delta=10^{-3},\ \mu=-0.99\Delta$]{%
  \includegraphics[width=8cm]{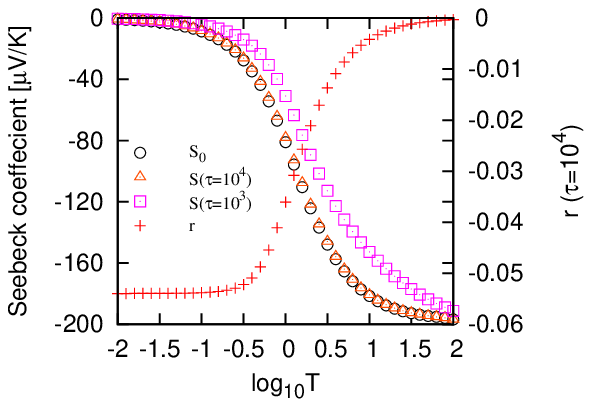}}%
 \subfigure[$\eta=2,\ \Delta=10^{-5}, \ \mu=0.99\Delta$]{%
  \includegraphics[width=8cm]{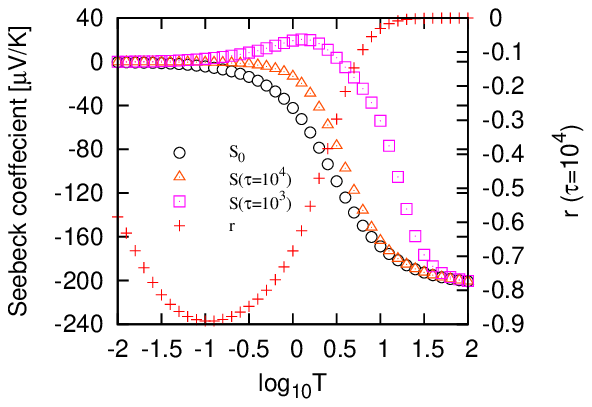}}%
 \caption{(Color online) Temperature dependence of Seebeck coefficients (left axis) $S_0$ (circle), $S$ for $\tau=10^4$ (square, pink) and $\tau=10^3$ (rotated square, blue)  and of the relative strength of anomaly $r$ (right axis, plus sign, red) for $\tau=10^4$. }  \label{S-T}
  \end{figure}
\end{center}

Fig.\ref{S-T} (a) is a plot for the case of $\eta=1$ (the special point as pointed out above), $\Delta=10^3$, and the chemical potential $\mu$ situated only one  hundredth of $\Delta$ above the $\varepsilon_-$ band edge.  One cannot distinguish, on the plotted scale,  $S$ for $\tau=10^5$ (rather clean case, not shown in figure) from $S_0$.  For 10 times dirtier case ($\tau=10^4$), the absolute value of $r$, by which we evaluate the extent of anomaly, reaches $\simeq$ 4\% around $T \simeq 1{\rm K}$.  Further increase in the scattering strength ($\tau=10^3$) makes the anomaly much larger in a wider range of $T$.  

Fig.\ref{S-T} (b) is plotted for $\eta=2$ (of the same order as Fig.\ref{S-T} (a)),  100 times smaller exchange $\Delta$ compared to Fig.\ref{S-T} (a), and $\mu$ put at $\mu=0.99\Delta$.  The general dependence on $\tau$ is similar to that in Fig.\ref{S-T} (a), except that  for $\tau=10^3$, $S$ varies non-monotonically with $T$ and even shows hole-like sign $S>0$ in lower  $T$ range.

Regarding both (a) and (b) in Fig.\ref{S-T}, one recognizes that in all cases (considering Berry curvature or not, and independent of the value of $\tau$), Seebeck coefficient approaches to zero as $T$ is lowered, in consistent with well-known Mott's law.  On the high $T$ side, $S$ values are converging into $S_0$, which is because $\sigma_{xx}$ and $\alpha_{xx}$ increase as a function of $T$ at least at a faster rate than their transverse counterparts (The latter ones even begin to decay at some $T$ in the plotted range as $T$ increases in the case of Fig.\ref{S-T} (b).  The reason is that, due to its small split $\Delta \approx k_{\rm B}T$ at $T  \approx 1{\rm K}$, smoothed function $f(\varepsilon)$ suppresses $\sigma_{xy}$ by causing  cancellation of $\pm \Omega$ on each band and that broadened function  $\partial f/\partial \varepsilon$ inhibits the grow of $\alpha_{xy}$.)

The above mentioned trend regarding $\tau$  is predictable from the explicit relations $\beta \propto \tau^{-1},\ \gamma \propto \tau^{-1}$(see the definitions in Section 2.).    However, we have to be
careful  regarding this result, since shorter $\tau$ means worse defined
quantum number {\bf k}.  There is also a possibility that it is
necessary to take into account extrinsic
scattering-related mechanisms of AHE and ANE by applying 
techniques such as  intuitive semiclassical theories(that include many
other contributions than ours with only the intrinsic one) or more
rigorous quantum theories(Kubo-Streda formula or Keldysh formalism).

\section{Summary}
The semiclassical formula for Seebeck coefficient that includes the contribution of  Berry curvature was derived.  
  For a special case of Zeeman-Rashba 2DEG, the following result was obtained: When (i) the temperature is rather low, (ii) scattering is not too weak, and (iii) the band shape has some specific feature, Berry curvature  could have quite large effect on Seebeck coefficient. It is to be studied to what extent extrinsic AHE and ANE
 modify our results.   
 
 \section*{Acknowledgements}
The authors thank the Yukawa Institute for Theoretical Physics at Kyoto University.
Discussions during the YITP workshop YITP-W-13-01 on
"Dirac electrons in solids" were useful to complete this work.
Part of this research has been funded by the MEXT HPCI Strategic Program.
This work was partly supported by Grants-in-Aid for Scientific Research
(Nos. 25104714, 25790007, and 25390008) from the JSPS.




\end{document}